\documentclass[twocolumn,showpacs,preprintnumbers,amsmath,amssymb,floatfix]{revtex4}

\usepackage{pst-all}
\usepackage{graphicx}
\usepackage{color}
  \definecolor{green}{rgb}{0.0,0.6,0.0}
  \definecolor{orange}{cmyk}{0.0,0.5,1.0,0.0}

\usepackage{dcolumn}
\usepackage{bm}
\begin{document}
\preprint{BNL-81489-2008-JA}

\title{RF Breakdown with and without External Magnetic Fields }
\author{R. B. Palmer, R. C. Fernow, Juan C. Gallardo, Diktys Stratakis}
 \affiliation{Brookhaven National Laboratory, Upton NY 11973}
\author{Derun Li}%
\affiliation{Lawrence Berkeley National Laboratory, Berkeley, CA 94720}%
\date{\today}

\begin{abstract}
Neutrino Factories and Muon Colliders' cooling lattices require both high gradient rf and strong focusing solenoids. Experiments have shown that there may be serious problems operating rf in the required magnetic fields. The use of high pressure gas to avoid these problems is discussed, including possible loss problems from electron and ion production by the passage of an ionizing beam. It is also noted that high pressure gas cannot be used in later stages of cooling for a muon collider.   Experimental observations using vacuum rf cavities in magnetic fields are discussed, current published models of  breakdown with and without magnetic fields are summarized, and some of their predictions compared with observations.

A new theory of magnetic field dependent breakdown is presented. It is proposed that electrons emitted by field emission on asperities on one side of a cavity are focused by the magnetic field to the other side where they melt the cavity surface in small spots. Metal is then electrostatically drawn from the molten spots, becomes vaporized and ionized by field emission from the remaining damage and cause breakdown. The theory is fitted to existing 805~MHz data and predictions are made for performance at 201~MHz. The model predicts breakdown gradients significantly below those specified for either the International Scoping Study (ISS)~\cite{iss} Neutrino Factory or a Muon Collider~\cite{collider}.

Possible solutions to these problems are discussed, including designs for `magnetically insulated rf' in which the cavity walls are designed to be parallel to a chosen magnetic field contour line and consequently damage from field emission is suppressed.
An experimental program to study these problems and their possible solution is outlined.
\end{abstract}

\maketitle

\section{\label{sec:level1}Introduction}
\subsection{The use of rf in Magnetic Fields for Neutrino Factories and Muon Colliders}
Low frequency (330-200~MHz) rf is needed for phase rotation and early cooling in the currently  proposed
Neutrino Factory~\cite{iss} and Muon Collider~\cite{collider} designs. The magnitude of the required magnetic fields are of the order of 1.75~T for the phase rotation and around 3~T in the early cooling lattices. The rf gradients specified are between 12 and 15~MV/m~\cite{neuffer}. 

Experimental data exists on the operation of 805~MHz vacuum rf in magnetic fields. These were obtained with two very different cavity types: 
1) a multi-cell cavity with open irises~\cite{noremopen}, and 
2) a single `pillbox' cavity with irises closed with Cu plates or Be windows~\cite{xxx},\cite{norempill}. Both showed significant problems and will be discussed in this article.


There is very little data on operations of  201~MHz vacuum cavities in significant magnetic fields, although experiments are underway with local fields of about 1~T on a limited part of the cavity. Tests are planned for the 201~MHz cavity in fields of the order of 3~T and geometries that can be made close to those in the machine designs.\footnote{These tests will be undertaken when the first large superconducting `Coupling Coil' is delivered from Harbin, China, within the next 12 months.}

There is also data on the DC operation of a test cavity in high pressure hydrogen gas  that showed no magnetic field dependence~\cite{gas}. However, there may be other problems arising  when an ionizing beam passes through such a gas filled cavity; in addition, such gas filled cavities cannot be used in the later stages of cooling for a Muon Collider because the Coulomb scattering in the gas would cause too much emittance growth.

We first discuss the gas filled cavity experiments and the possible problems with their use with ionizing beams; then
we consider the experiments and models of vacuum cavities ( no gas in them) in magnetic fields.
\subsection{805~MHz High Pressure Gas filled Test Cavity Experiments} 
\begin{figure}[!ht]
\includegraphics[width=2.in]{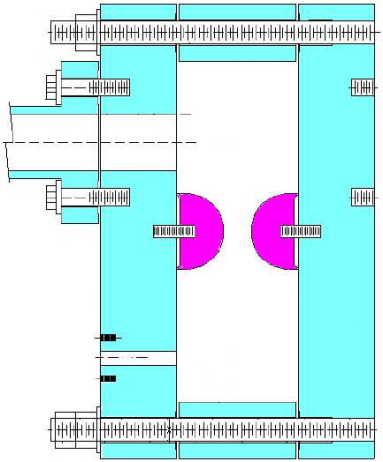}
\caption{\label{gas} (Color) Schematic of a test cavity.}
\end{figure}
A simple test cavity (Fig.~\ref{gas}) has been operated with hydrogen at different pressures and with `buttons' made of different materials  that define the small high gradient gap~\cite{gas}. It was found that at lower pressures, the breakdown gradient follows the Paschen prediction, but at higher pressures the gradient is limited to values (50~MV/m for Cu) close to those observed ($53~\frac{\text{MV}}{\text{m}}$ in the open multi-cell cavity discussed in Sec.~\ref{exp-multi}) in vacuum cavities after conditioning.
This similarity in the observed maximum gradients suggests that the initiating mechanism for breakdown in the high pressure gas, and vacuum cavities is the same. Models for this phenomena will be discussed in Sec.~\ref{vacnofield}. No change in breakdown was observed in the presence of an external magnetic field of 3~T. This is as would be expected from the mechanism for vacuum breakdown with magnetic fields that is described in Sec.~\ref{vacfield}.
\subsection{Theoretical Expectation for the Effects of an Ionizing Beam passing through a High Pressure Gas filled Cavity}
Tollestrup~\cite{alvin1} has studied the likely effects of a muon beam passing through gas filled rf cavities. It is concluded that
 the lifetime of the electrons and ions produced by the ionization of those beams are long compared with the likely duration of the muon beams. It is further concluded that such electrons  in the cavity will be driven backwards and forwards as the rf voltage oscillates, and that this will lead to heating of the gas and loss of the rf energy. The final $Q$ of three different cavities exposed to $10^{11}$ muons are estimated and given in Table~\ref{tab:table1}.   
\begin{table}[!ht]
\caption{\label{tab:table1} Final $Q$ of three cavities.}
\begin{ruledtabular}
\begin{tabular}{lccc}
\hline
Frequency (MHz)&400&800&1600\\
rf gradient (MV/m)&16&16&16\\
Final Q&647&325&163\\
\end{tabular}
\end{ruledtabular}
\end{table}
The final $Q$ for other cases can be approximately   estimated  by scaling from these numbers. This is done for three cases:
\begin{itemize}
\item The phase rotation and initial cooling for the ISS Neutrino Factory~\cite{iss}
\item The phase rotation and initial cooling for the low emittance muon collider parameters as presented at \textsl{The Neutrino Factory \& Muon Collider Collaboration} (NFMCC) 2008 meeting~\cite{mainpage},~\cite{alexahin1}. 
\item The phase rotation and initial cooling for the high emittance muon collider~\cite{alexahin1}.
\end{itemize}

\begin{table*}[!ht]
\caption{\label{tab:table2}Parameters for a Neutrino Factory and low and high emittance Muon Collider.}
\begin{ruledtabular}
\begin{tabular}{lccc}
\hline
&$\nu$ Factory&\multicolumn{2}{c}{Muon Collider}\\
&&\small{Low emittance}&\small{High emittance}\\
Frequency (MHz)&200&200&200\\
rf gradient (MV/m)&16&16&16\\
Repetition rate (Hz)&50&65&12\\
Wall power (GW) &4&3.6&3.2\\
10 GeV protons on target ($10^{12}$) &51 &(35) &(170) \\
Final muons per charge ($10^{12}$)&  & 10 $\times$ 0.1& 2  \\
Cooling transmission   (\%)& &30&7\\
Muons per proton& 0.33&(0.17)&(0.33)\\
Muons before cooling, $N$ ($10^{12}$)& 17  &6   &57\\
Final Q&7.6&22&2.3\\
\end{tabular}
\end{ruledtabular}
\end{table*}

Table~\ref{tab:table2} gives the needed parameters of the three cases.
For the two collider cases, the numbers of initial muons are estimated from the final number of muons, multiplied by two for the 2 charges, and divided by the quoted transmission. This is a somewhat optimistic assumption since the phase rotation and initial cooling will also see unwanted kaons, very high energy pions, protons and knock-ons from neutrons, that will add to the losses.
For the neutrino factory case the total number of muons is derived from the number of protons per pulse at 10~GeV together with MARS~\cite{mars} simulated  total pion production at this energy.  

We assume that the final $Q$ of the cavity  is given  by
\begin{equation}
Q=Q_o \times \frac{f_o}{f}\times \frac{N_o}{N}
\end{equation}
where $f$ is the cavity frequency, $N$ is the number of muons per pulse and the subscripted values are from the 400~MHz case in Ref.~\onlinecite{alvin1}.

Since $Q$ tells us the energy loss per period of the rf, the $Q=22$ for the low emittance case means that little rf is left after only three cycles. 
Yet for the phase rotation schemes under study, the muons are spread over 10 or more cycles prior to bunching and rotation.
Further down the cooling channel the numbers of muons are  less, but until and unless the bunches are merged, the rf must remain effective over 10 or more cycles, and the required $Q$ must be well above 60~\footnote{The introduction of other gasses to rapidly capture the electrons may avoid this problem, but there remain concerns that the presence of accumulating numbers of ions will cause trouble.}.
 In any case, it must be noted that high pressure gas cannot be used in the later cooling stages where the emittance is very small. In these cases the Courant Snyder $\beta_\perp$ must be very small where any material is introduced. This can be achieved in local areas, or inside very high field solenoids, but not over the lengthy rf systems. Final cooling for a muon collider will thus inevitably require vacuum acceleration near strong fields.
\subsection{Experimental Operations of a Multi-cell 805~MHz Open Cavity in approximately Axial Magnetic Fields}
\label{exp-multi}
\begin{figure*}
\includegraphics[width=5.in]{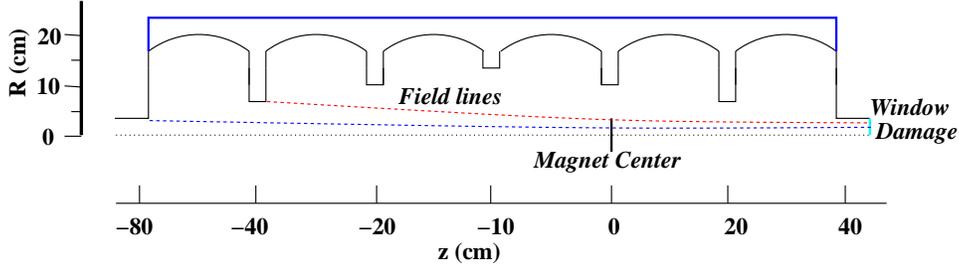}
\caption{\label{multicell} (Color) Schematic of a multi-cell cavity first tested in the lab G magnet at Fermilab. The lines shown are the magnetic field lines extending from two irises to the end window.}
\end{figure*}

An ionization cooling lattice had been designed that employed liquid hydrogen absorbers inside high field solenoids separated by lower field matching solenoids surrounding  6 cell, open iris, rf cavities. Such lattices used early in a cooling sequence, would  use 201~MHz, but higher frequency and high magnetic fields would be used for later cooling.

To test the system a 6 cell cavity (Fig.~\ref{multicell}) was designed with iris apertures tailored to fit the beam profile in the matching fields~\cite{noremopen}. The cavity was then tested with and without magnetic fields. The  available fields  were not those in the matching design, and the cavity could not be positioned symmetrically in the magnet, so the fields used in the test were asymmetrical. 
 The maximum achieved surface gradients were quite high ($\approx 53$~MV/m), comparable with those in the gas filled cavity, and were not strongly dependent on the magnetic fields (Fig.~\ref{pill}a). However, both X-rays and dark currents were greatly increased with the magnetic field on. After some time, vacuum was lost, and it was found that the end vacuum window was so damaged as to cause the vacuum leak.

It was found that the location of the window damage corresponded to a focused dark current coming from one of the high field irises. In addition radiation damage patterns were observed showing that many beamlets were  focused by the magnetic fields onto the end window. 
There was no indication that dark currents from one high gradient iris were focused on to another, which may explain why the maximum achieved gradients remained high (see Sec.~\ref{withmag}).

\subsection{  Pillbox Cavity Breakdown in Magnetic Fields}
\label{vacfield}
\begin{figure*}[!ht]
\includegraphics[width=2.5in]{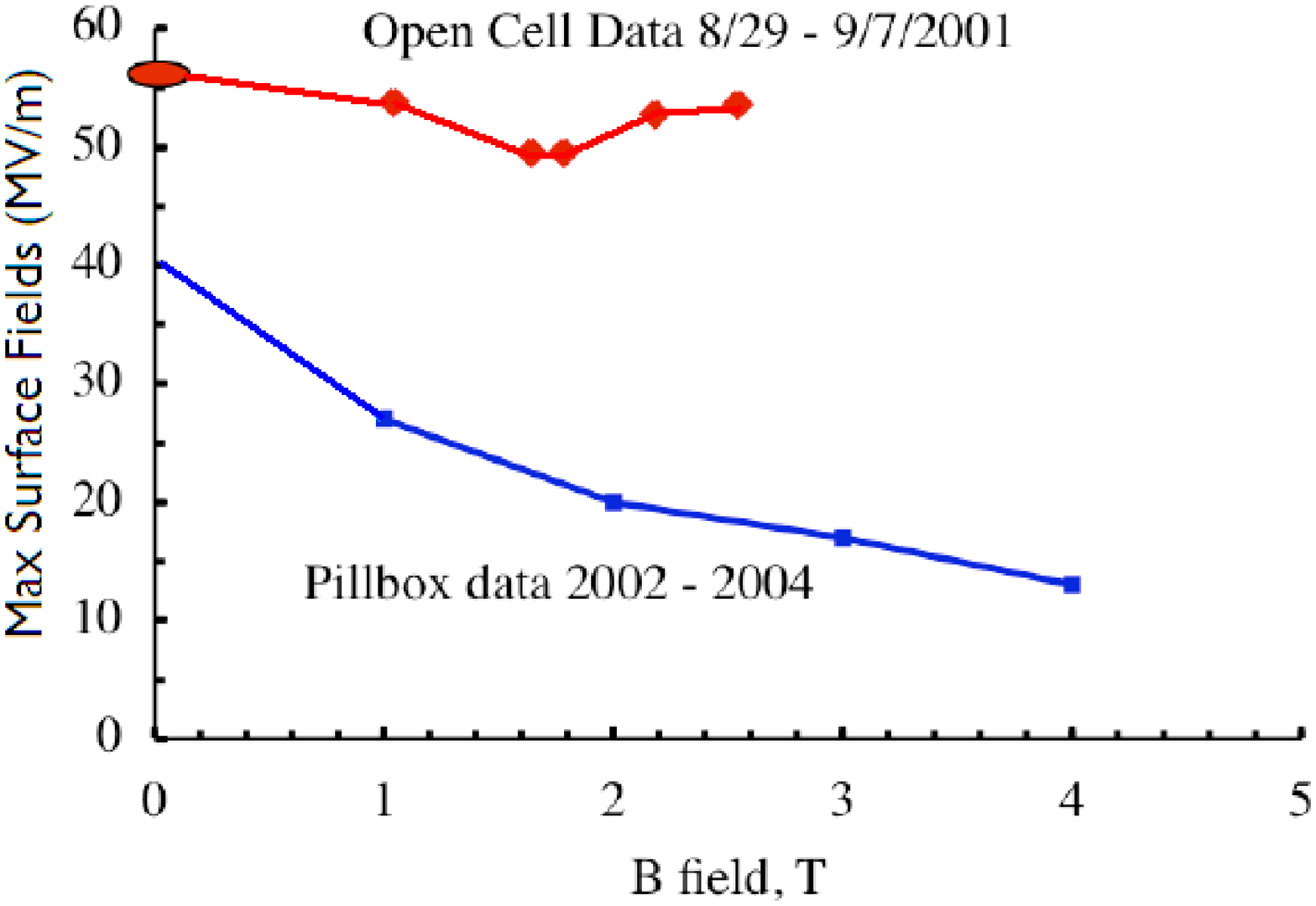}
\includegraphics[width=1.5in]{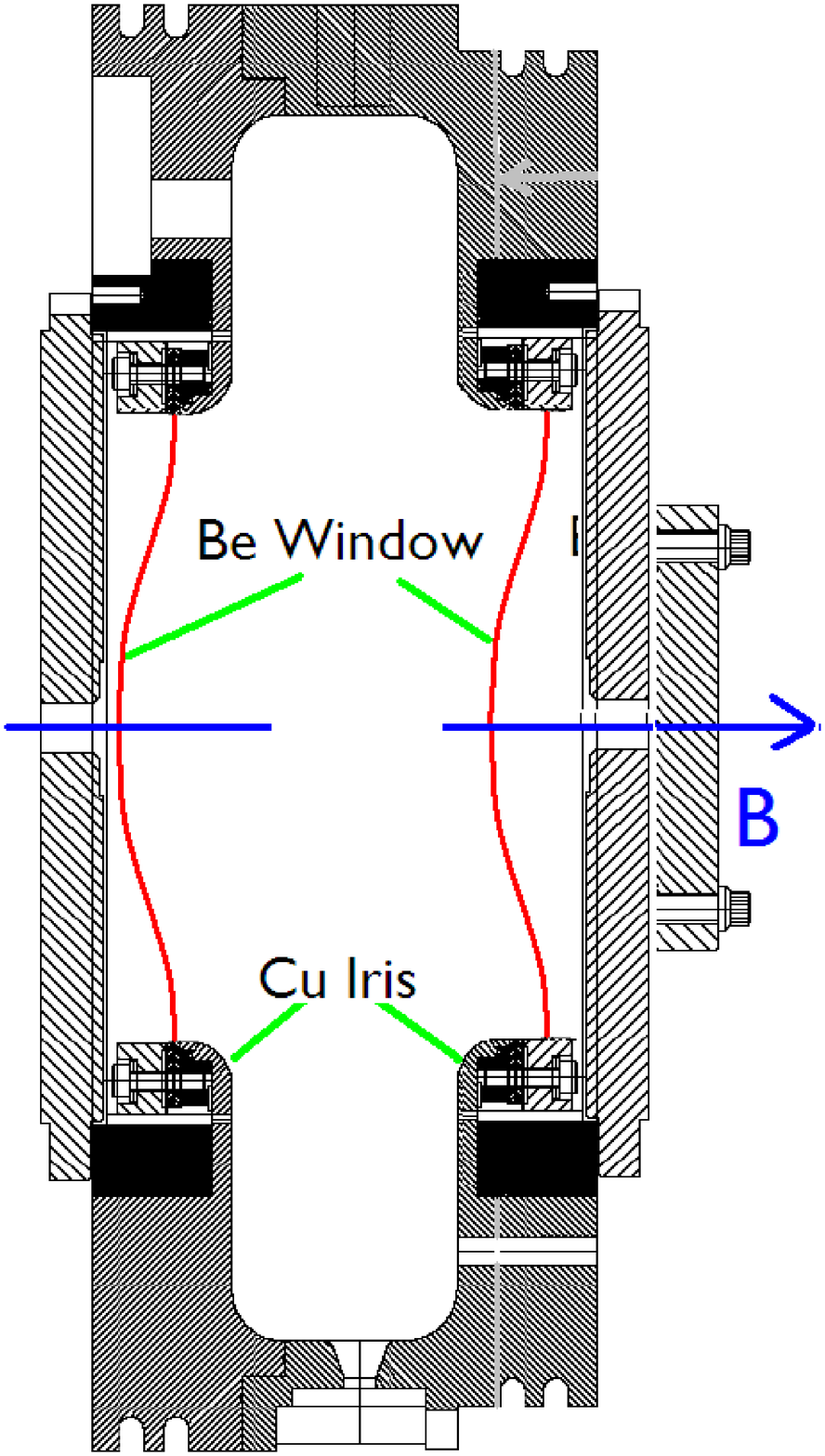}
\caption{\label{pill}(Color) Left: a) breakdown gradients \textit{vs} axial magnetic field. Right: b) schematic of a pillbox cavity. (Data from Ref.~\onlinecite{xxx})}
\end{figure*}

In a linac with open irises, the peak surface fields are typically a factor of two higher than the average accelerating gradient.
Subsequent to the design of the multi-cell open iris cavity, it was realized that with muons one could introduce thin Be windows at each iris and obtain accelerating gradients much closer to the maximum surface fields. In addition, the resulting `pillbox' cavities give more acceleration for a given rf power. A test `pillbox' cavity was thus designed (Fig.~\ref{pill}b) and tested in a magnetic field. This time the cavity was mounted  in the center of the magnet and could thus be tested with 
symmetrical fields. It was also tested with one of the magnets coils unpowered to give asymmetrical fields and with the coils powered in opposite directions to study the effects with these different field geometries. 

Without field the maximum achieved gradients~\cite{norempill} were somewhat lower than for the multi-cell cavity, but this was, at least to some extent, the result of more conservative operation after the severe damage seen in the earlier cavity. With magnetic fields, the maximum gradients were found to be strongly dependent on field (Fig.~\ref{pill}a). It was also found that over time its performance deteriorated. Examination of the inside of the cavity showed severe pitting on the irises. The Be windows themselves showed no visible damage, but there was a  spray of Cu over their surface and Cu powder in the bottom~\cite{splash}.
\section{\label{vacnofield}Breakdown models without magnetic fields}

It is assumed in all models that breakdown is initiated at `asperities', where the local electric fields is higher, by a factor, $\beta_{\rm FN}$ introduced by Fowler-Nordheim~\cite{refFN}. The average values of these factors can be determined by observing the electron currents (dark current) emitted by the sum of many asperities, each of which has a specific value of $\beta_{\rm FN}$.
The field emitted average electron current density $J_F(\frac{\text{A}}{\text{m}^2})$ for a surface field $\bm{E}(\frac{\text{V}}{\text{m}})$, and local field $\bm{E}_{\text{local}}=\beta_{\rm FN}\bm{E}$ is given by~\cite{ohmic}
\begin{equation}
J_F =6\times 10^{-12}\times 10^{4.52\phi^{-0.5}}\frac{\bm{E}_{\text{local}}^{2.5}}{\phi^{1.75}}
\exp{\left[-\frac{\zeta \phi^{1.5}} {\bm{E}_{\text{local}}}\right]}
\label{eq1}
\end{equation}
where $\phi$ is the material work function in (eV) ($\phi=4.5$\,eV for Cu) and 
\begin{equation}
\zeta=6.53\times 10^9 (\text{eV})^{-1.5} (\frac{\text{V}}{\text{m}}).
\label{eq11}
\end{equation}

In vacuum cavities with a thin window, one can measure some fraction of all field emitted electrons and observe its field dependence. From this dependence, given an assumed  work function $\phi$, one can then extract an average value of $\left<\beta_{\rm FN}\right> \left<\bm{E}\right>$.

It is reasonable to assume that breakdown occurs where the local field, and thus $\beta_{\rm FN}\bm{E}$ is maximum, and thus higher by a factor $\alpha$  than the average value determined from the gradient dependence of the dark current from many asperities, so
\begin{equation}
\frac{\bm{E}_{\rm local}}{\alpha} =\left<\beta_{\text{FN}}\right> \left<\bm{E}\right>
\end{equation}
where $\alpha \agt 1$ depends on the probability distribution of  $\beta_{\rm FN}$.

Observed breakdown gradients are found to depend on frequency~\cite{ref:freq} ($\propto \sqrt{f}$), rf pulse length, and cavity dimensions, but it has been found~\cite{noremfreq} that, over a range of frequencies from DC to a few GHz, and for differing pulse lengths, cavity dimensions and in waveguides, the values of $\frac{\bm{E}_{\text{local}}}{\alpha}$ fall in a relatively narrow range around 7~GV/m. Breakdown thus appears to be related to the local electric fields at asperities (points, cracks or other causes of the field enhancement), or to the field emitted currents that are strongly dependent on these fields.
Several plausible but speculative mechanisms for the initiation of rf breakdown have been proposed, which will be discussed next.
\subsection{Breakdown Models}
\subsubsection{Mechanical fracture model}
 The mechanism, as discussed in Ref.~\onlinecite{noremmodel}, assumes the following sequence of events:

1) The surface contains asperities at the top of which the local field is given by the average field multiplied by a Fowler-Nordheim field enhancement factor $\beta_{\text{FN}}$. 
2) The outward electrostatic tension $F_s=\frac{\epsilon_o}{2}(\beta_{\text{FN}}\bm{E})^2$ is equal in magnitude to the energy density of the field, hence it is proportional to the square of the field (7~GV/m would induce a tension of 300~MPa), breaks off the tip  and the small piece now moves away from the remainder of the asperity.
3) The piece is bombarded by field emitted electrons from the remaining asperity and becomes vaporized and ionized.
4) Following this formation of a local plasma
other mechanisms cause the plasma to spread, or other mechanisms short out the cavity leading to breakdown.

\begin{figure}[!ht]
\includegraphics[bb=12 490 476 762,clip,width=2.75in]{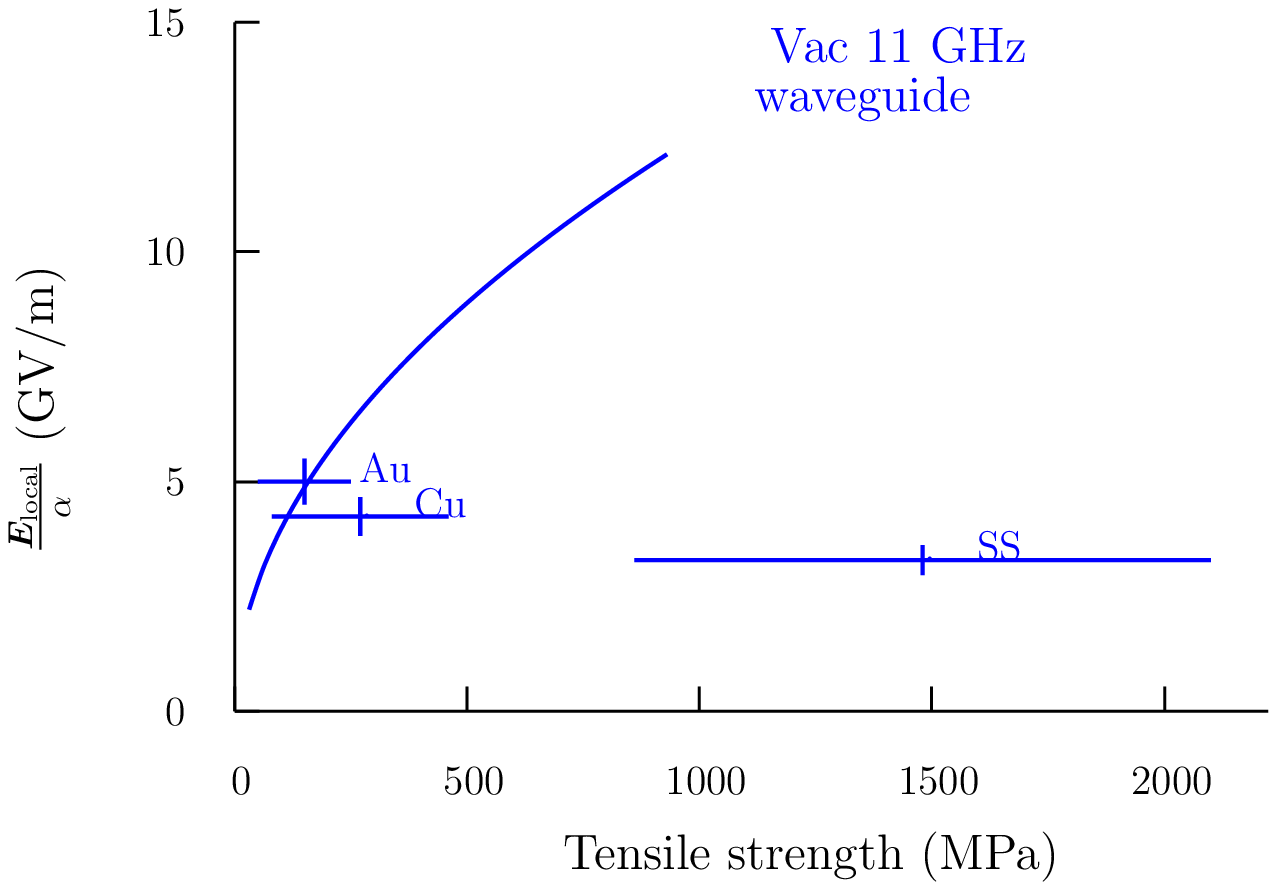}
\includegraphics[bb=30 490 470 750,clip,width=2.75in]{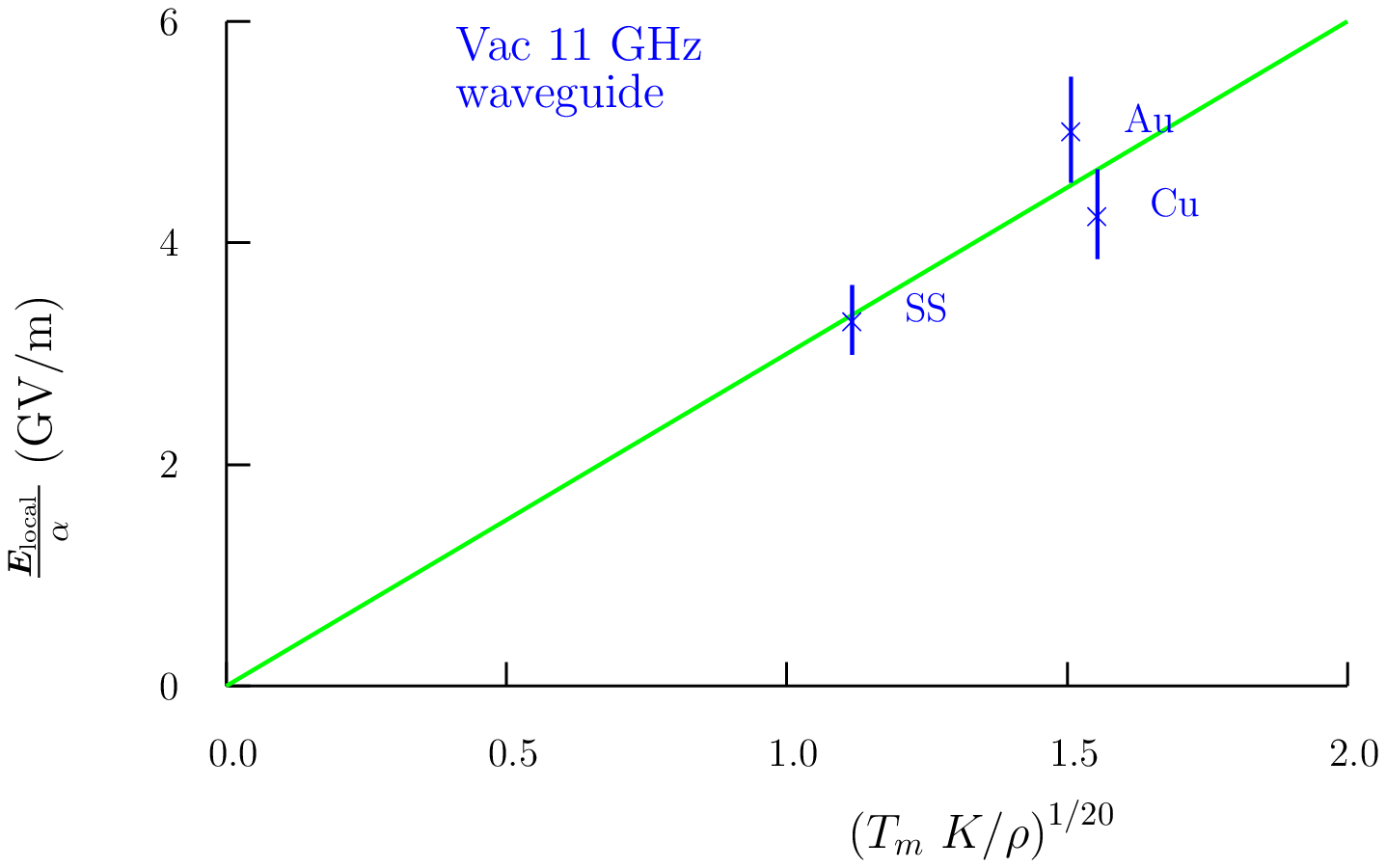}
\caption{\label{model1}(Color) Top: a) minimum local rf gradient at an asperity over $\alpha$ \textsl{vs} tensile strength of materials of an 11~GHz waveguide; the error bars on tensile strengths indicate the range of quoted values, depending on material treatment. Bottom: b) local rf gradients \textsl{vs} the product of the melting temperature $T_{\text {m}}$ times the thermal conductivity $K$ divided by the electrical resistivity $\rho$ for an 11~GHz vacuum waveguide. (Data from Refs.~\onlinecite{gas},~\onlinecite{wg})}
\end{figure}

Within this model the breakdown occurs when the electrostatic outward tension at the asperity equals the tensile strength of the material. For the local field $\bm{E}_{\text{local}}=\alpha \left<\beta_{\text{FN}}\right> \left<\bm{E}\right>$, the average field at the surface of the asperity is given by
\begin{equation}
\left<\bm{E}\right>\propto \frac{\sqrt{\textbf{T}}}{\left<\beta_{\text{FN}}\right>\alpha}
\end{equation}
where $\textbf{T}$ is the tensile strength of the material. There is relatively little data on breakdown, including $\beta_{\rm FN},$ of materials of significantly different tensile strength. Figure~\ref{model1}a shows the observed dependencies in an experiment~\cite{wg} at 11~GHz in a special tapered rf waveguide. The error bars on tensile strengths indicate the range of quoted values, depending on material treatment, with the highest strengths given for fine drawn wire.  It is reasonable to assume that on the nanometer scale of an asperity, the material strength is of the order of that the highest strengths observed.  The plotted curve represents the theoretic predictions assuming $\alpha=1$. However, independently of the value of $\alpha$, the theoretical expectation is not observed in the experimental dependency; indeed, the higher strength stainless steel has a lower minimum local breakdown field than expected.
\subsubsection{Ohmic heating model}
 It has  been suggested~\cite{ohmic} that, when the field emission current density is sufficiently high,  breakdown is initiated by ohmic heating that  melts the tip of an asperity. Once liquefied electrostatic forces would pull the molten material away, just as the broken piece of the asperity was pulled away in the first model. This molten material, as it lifts from the remains of the asperity, will be exposed to field emission from the remaining asperity left behind and it will be further heated, vaporized and ionized to form a plasma. For submicron asperities, the time constant for achieving a steady thermal state is only of the order of a nanosecond, so the temperatures reached depend only on the geometry, electrical resistivity, thermal conductivity, and current densities at the tip.
\begin{figure}[!ht]
\includegraphics[width=1.in]{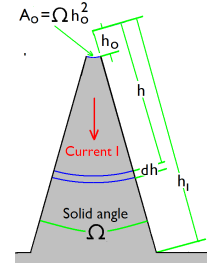}
\caption{\label{asperity}(Color) Schematic for asperity heating calculation}
\end{figure}
Assuming  the asperity to have a conical shape (see Fig.~\ref{asperity}), with solid angle $\Omega$, and emitting area $A=h_o^2\Omega,$ then, given the electrical resistivity $\rho$ and current density $j_o$, the heat $Q$ flowing back to the base of the asperity is
\begin{equation}
Q(h)= \int_{{h_0}}^{h}\,\frac{I^2~\rho}{\Omega h^2}\,dh\approx\frac{ I^2~\rho}{\Omega}\left( \frac{1}{h_0}-\frac{1}{h}\right)
\end{equation}
The temperature difference between the tip and the base $\Delta T$, as a function of the thermal conductivity $K$, assuming $h_1 \gg h_o$ is
\begin{eqnarray}
\Delta T = \int_{{h_0}}^{{h_1}}\, \frac{Q}{Kh^2\Omega}\,dh &\approx &\left(\frac{I^2 \rho}{2h_0^2K\Omega^2} \right)\nonumber \\
&=&\left(\frac{ j_o^2A\rho}{2K\Omega}\right)
\end{eqnarray}
The field emission current density  is approximately proportional to the local field
 $\bm{E}_{\text{local}}$ to the tenth power~\cite{noremopen}, so for a fixed emission area $A$ and cone angle $\Omega$, the field needed to melt the material is proportional to $\bm{E}_{\text{local}}\propto \left(\frac{K~T_m}{\rho}\right)^{1/20}.$ This expression contains a number of approximations: the temperature dependence of the parameters is ignored, the asperity shape is assumed to be a cone of fixed angle, and the field emission is approximated by a power law, but the result should be qualitatively correct, and is plotted in Fig.~\ref{model1}b. It is seen  that the  material dependencies agree with the prediction within the errors, and, in particular, the  lower gradients achieved with stainless steel are as predicted.
\subsubsection{Thermal runaway model}
\begin{figure}[!ht]
\includegraphics[bb=80 480 484 762,clip,width=2.75in]{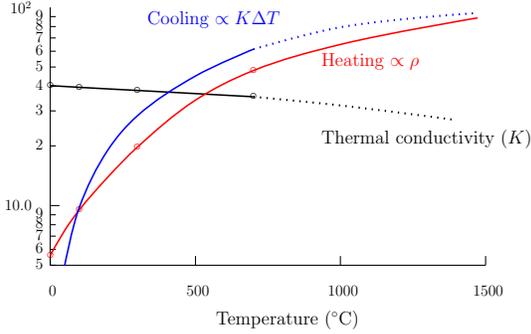}
\caption{\label{vstemp}(Color) Rates of heating from electrical resistivity $\rho$, and cooling from thermal conductivity $K$, \textsl{vs} the effective temperature at which $\rho$ and $K$ are determined.}
\end{figure}
The above calculation  of ohmic heating was carried out under the assumption that the resistivity and thermal conductivity are independent of temperature, which they are not. Assuming that these expressions are approximately valid using resistivities and thermal conductivities  at an intermediate effective temperature $T_{\text{effective}}$, then one can look at the relative rates of heating and cooling as a function of that effective temperature. Figure~\ref{vstemp} shows the heating and cooling vs the effective temperature for pure Cu~\cite{kandl}. For a low current density $j_0$  giving a temperature less than a critical value $(T_{\text{effective}} < T_{\text{critical}}),$ then  the cooling which is $\propto \Delta T \times K$ rises more rapidly than the  heating $(\propto \rho(T))$, and a stable temperature is possible. On the other hand at higher current densities, which leads to temperatures above $T_{\text{critical}}$, the rate of heating vs temperature rises faster than the  cooling and the temperature will `runaway'. For the data used, the critical effective temperature is about $300^{\circ}.$
 The actual critical temperature at the tip will be somewhat higher. It is worth noting that repeated heating to temperatures of this order may induce fatigue, leading to damage and an increased probability of breakdown after many rf pulses.
\subsubsection{Reverse bombardment model}
 A fourth model~\cite{wilson} assumes that some initial mechanism generates a local plasma (called a `plasma spot'), that by itself does not directly cause breakdown. Plasma spots, tiny sources of light, have been observed in DC and pulsed gaps without breakdown. In this model, breakdown occurs when electrons emitted by the local plasma are returned to their source spots by the rf electric field. The energy given to the source by these returning electrons is required to cause the plasma to grow and cause the actual breakdown. But as we will see later (Fig.~\ref{cavel}), at least for 805 and 201~MHz  in the absence of an axial magnetic field, electrons emitted at the highest field location never come back to their source, no matter what their initial phase. This appears to be true for cavities in general, with the exception of emission on the axis in pillbox cavities. Yet breakdown on the axis of pillbox cavities is rarely observed. 
\subsubsection{Surface damage by heating}
As we have noted, breakdown gradients rise approximately as the root of the rf frequencies; but this dependency does not continue at frequencies above 10~GHz. Fatigue damage from  cyclical surface heating appears then to limit the gradients. This damage is worse at locations with maximum surface currents, where surface electric fields are usually low. But the damage, which causes cracks to form at grain boundaries can be so severe that breakdown is initiated in regions with both electric fields and surface current. Since this phenomena is only seen at such high frequencies, it is not expected to be a problem for neutrino factories or muon colliders.
\subsection{Observed Dependencies}
\begin{figure}
\includegraphics[bb=10 493 480 763,clip,width=3.in]{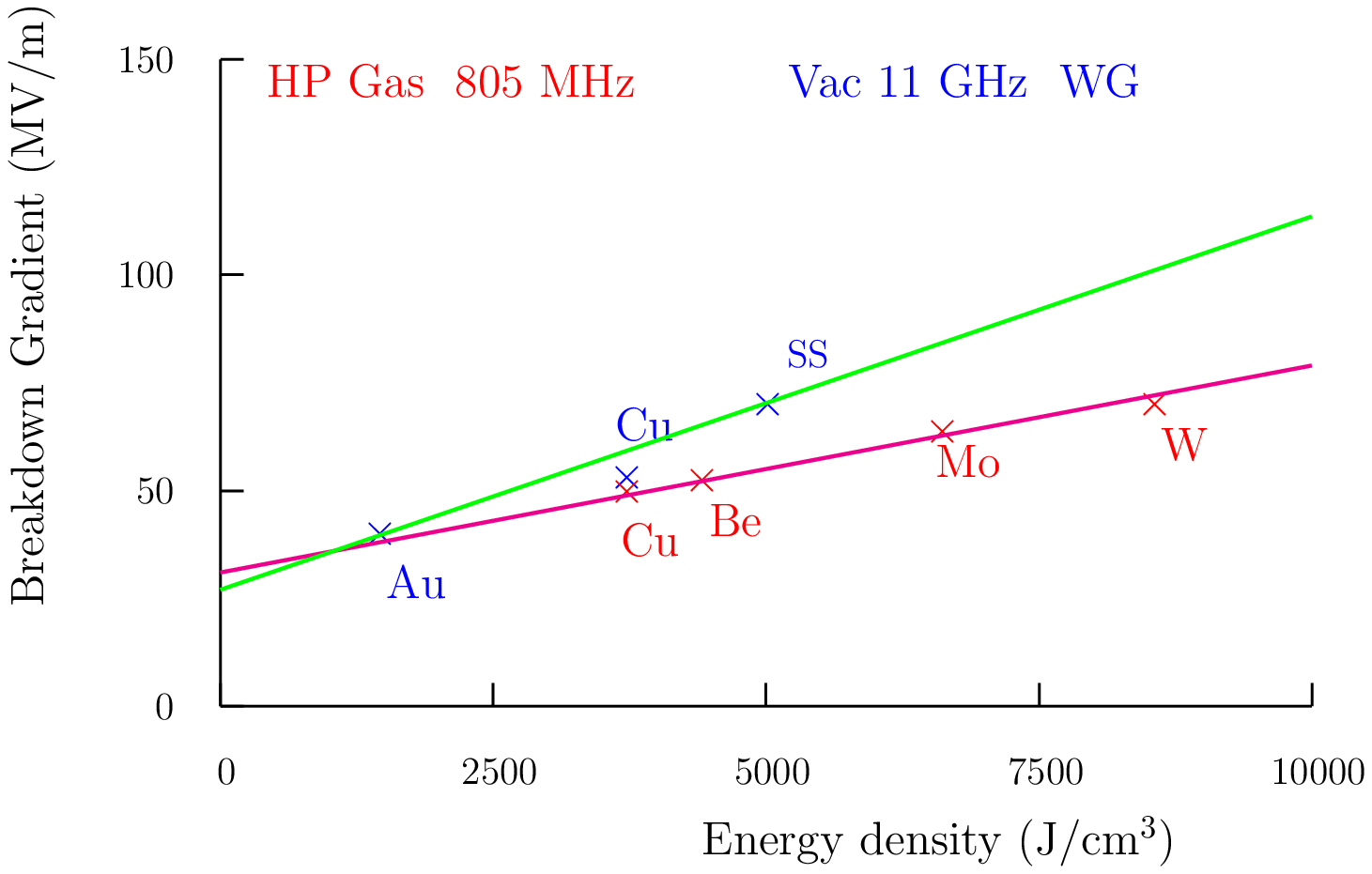}
\includegraphics[bb=30 493 460 756,clip,width=3.in]{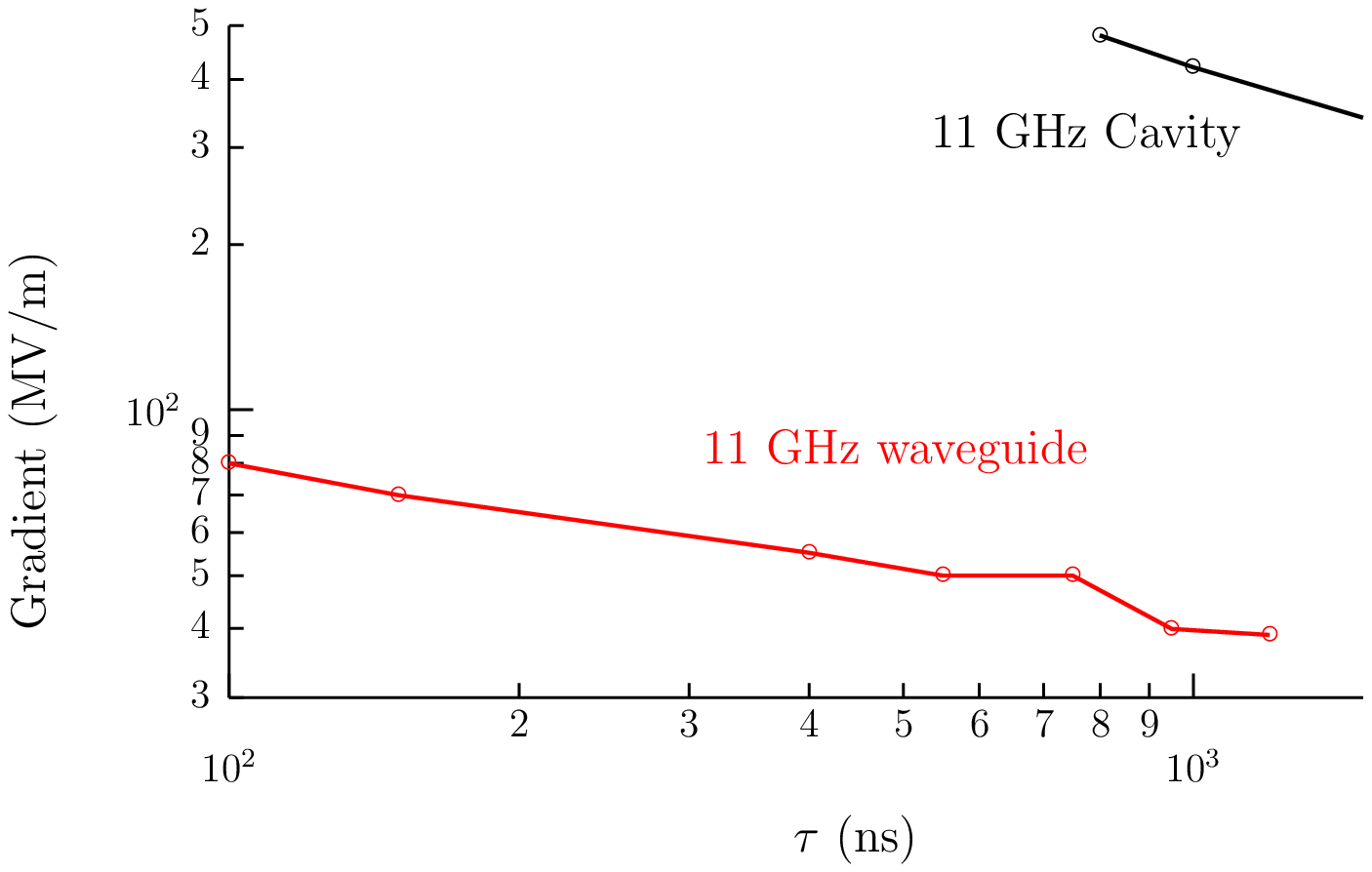}
\caption{\label{model2}(Color) Top: a) breakdown gradients \textit{vs} the approximate energy to melt a given volume of the material. Bottom: b) breakdown gradients \textit{vs} the rf pulse length. (Data from Refs.~\onlinecite{gas},~\onlinecite{wg})    }
\end{figure}
Although the above discussion suggest that ohmic heating initiates breakdown, this conclusion depends on a limited number of experiments; therefore one has to conclude that the initial mechanism that starts a breakdown is not fully understood yet. But there are some dependencies that appear fairly consistently:
\begin{itemize}
\item Over a wide range of frequencies (0.2 to 3~GHz), breakdown gradients are approximately proportional to the square root of the rf frequency, and this dependency arises because of changes in the observed field enhancement factor $\beta_{\text{FN}}$.
\item Breakdown appears dependent on the required energy to melt a given volume of the electrode material (see Fig.~\ref{model2}a). In the vacuum waveguide experiment case, this dependence is again caused by changes in $\beta_{\text{FN}}$, rather than in the local fields.
\item Breakdown occurs at lower gradients for long rf pulses than for short pulses (Fig.~\ref{model2}b), and there is some indication that this too arises from the pulse lengths influence on $\beta_{\text{FN}}$.
\end{itemize}
\subsection{Conditioning}
 In all cases discussed here, the cavities or waveguides were `conditioned' prior to achieving the quoted gradients. After one or more breakdowns at one gradient, the cavity will subsequently withstand a somewhat higher gradient. Typically, many hundred successive breakdowns are induced prior to the cavity reaching its `final gradient'. When dark current distributions are studied as the gradient is increased by this conditioning, then  it is again found~\cite{conditioning} that it is the enhancement $\beta_{\text{FN}}$ that is decreasing and not that the local field is increasing.

A reasonable assumption is that, whatever the initial cause, the effect of a breakdown is to remove asperities and consequently lowering $\beta_{\text{FN}}.$ At the same time, however, the breakdowns, depending on the energy available, will create new asperities thus incrasing $\beta_{\text{FN}}.$ In a `conditioned' cavity, these competing processes have reached an equilibrium. The rf energy available (stored or in a longer pulse), divided by the energy required to melt a given volume of metal, gives the total amount of melted metal. Assuming that more molten metal will be more likely to create bad asperities, leads directly to the observed dependencies with material properties and rf pulse length. 

In the case of the frequency dependency, since lower frequency cavities are usually larger, the observed dependency once again agrees with the expectation.
\section{Breakdown Models with External Magnetic Fields}
\subsection{Published Breakdown Model with Magnetic Fields}
It has been proposed in the twist model~\cite{noremopen} that the magnetic field dependence on breakdown arises from the torque forces on an asperity due to the inflow of current feeding the field emission reacting to the external magnetic field; that is $ F~\propto~I~\times~B$
 where $\bm{E}$ is the cavity electric field gradient and approximately $I~\propto~\bm{E}^{10}$~\cite{noremopen}.  
Thus for breakdown at a fixed force $F$, we expect
\begin{equation}
\bm{E}_{\text{breakdown}}~\propto B^{-1/10}
\end{equation}

\begin{figure}
\includegraphics[bb=14 475 479 764,clip,width=3.0in]{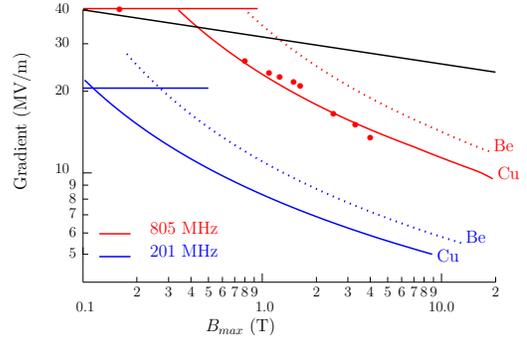}
\caption{\label{fit}(Color) Breakdown gradients \textit{vs} axial magnetic fields. Black line is dependency predicted by asperity twist model. Red lines are fit to the plotted Lab G~\cite{norempill} breakdown data. Blue lines are the calculated values for a 201~MHz cavity. Dotted lines are for Be surfaces.}
\end{figure}
In Fig.~\ref{fit} the observed pillbox cavity breakdowns are plotted as a function of the external average magnetic field.
The points plotted are those where both superconducting coils were powered so that the magnetic fields were relatively uniform over the cavity. 
The dependency predicted by this mechanism is shown by the solid black line; it is a poor fit at higher fields where the breakdown gradient falls much faster than predicted.
\subsection{Introduction to the Proposed Mechanism}
\label{withmag}
\begin{figure}
\includegraphics[width=2.5in]{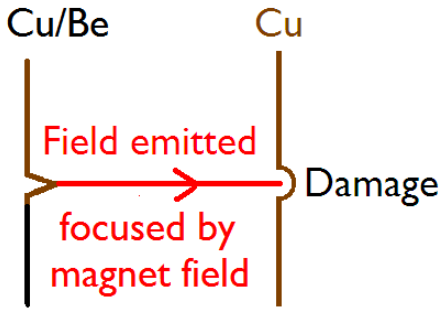}
\caption{\label{magrf}(Color) Proposed mechanism for breakdown with an external magnetic field.}
\end{figure}
We propose a new model  for breakdown with a magnetic field that is independent of  the breakdown mechanism  in the absence of magnetic fields (see Fig.~\ref{magrf}). Breakdown occurs by this mechanism \textsl{only} if its breakdown gradient is lower than  that from the case without a magnetic field. Its elements are:
\begin{itemize}
\item `Dark Current'  electrons are field emitted from an asperity,   accelerated  by the rf fields, and impact another location in the cavity. In the absence of a magnetic field these impacts are spread over large areas and do no harm.
\item With sufficient magnetic field they are focused to small spots, where they can melt the surface producing local damage. If such damage is at a low gradient location there is no immediate breakdown, but the damage can accumulate until, for instance, a hole is made in a window.
\item If the  electrons are focused onto a location  with high surface rf gradient, then electrostatic forces will pull the molten  metal out and away from the surface. This metal, as it leaves the now damaged location, will be exposed to field emitted electrons from the damaged area and will be vaporized and ionized, leading to a local plasma and subsequent breakdown.
\item For higher energy electrons, the  melting will start deeper in the material where the ionization loss is greater, and expand to the surface. Thus, when the melting reaches the surface, significant quantities of molten metal can  be sprayed onto other surfaces in the cavity~\cite{splash}. 
\end{itemize}
Breakdown will be dependent on a) the Fowler-Nordheim field enhancement $\beta_{\text{FN}}$ that determines the strength of the field emitted current, b) the local geometry of the asperity that will determine the initial particle distributions and effects of space charge, and c) on the geometry and magnetic fields that focus the electrons onto other locations.
\subsection{Electron Motion in a Cavity}
\begin{figure*}[!th]
\includegraphics[width=5.5in]{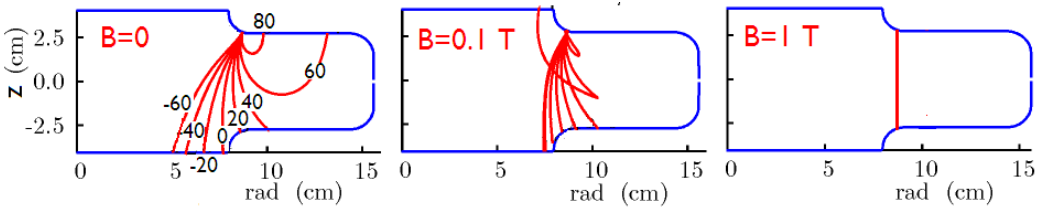}
\caption{\label{cavel}(Color) Trajectories of electrons field emitted at different phases from the highest surface field location in an 805 MHz pillbox cavity with a) no external magnetic field, b) an axial field of 0.1 T, and c) an axial field of 1 T. The axial electric field is 25 MV/m. Phases are in degrees relative to the maximum.  }
\end{figure*}
A  program CAVEL~\cite{cavel} tracks particles from arbitrary positions on the walls of a cavity until they end on some other surface. The program uses SUPERFISH~\cite{superfish} to determine the rf electric and magnetic fields, and uses a map of external magnetic fields calculated for arbitrary coil dimensions and currents.
Fig.~\ref{cavel}  shows  trajectories, for differing initial rf phases, starting from the highest field location on an iris. Without an external magnetic field, none of the trajectories from the high field location come back to their common origin. Tracks emitted at a phase of $20^{\circ}$  do hit the opposing iris at a high gradient location, but they are not focused there and  are spread out over a significant distance. But with a sufficient external axial magnetic field, the tracks are either focused to the high gradient location on the opposite iris or returned to their source. 

\begin{figure}[!h]
\includegraphics[width=3.in]{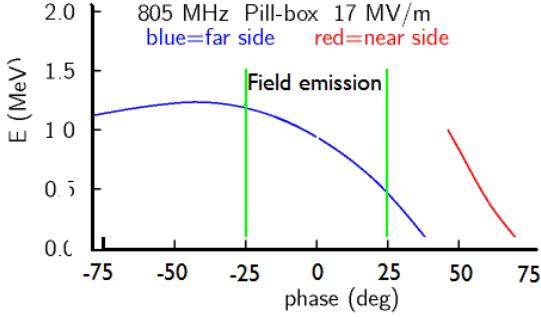}
\caption{\label{vsphase}(Color) Energies of electrons on impact vs. their phase of emission. Red indicates electrons that returned, blue those that  arrive on the next iris. Axial gradient is 17~MV/m. }
\end{figure}

There remains a small dependency of the arrival positions with phase, which arises from the combined effects of the perpendicular external and rf magnetic fields, but this dependency is small. If there were no other mechanism to spread out the electrons, then damage would appear as lines, which are not observed. In addition, damage would be much worse on the axis than at larger radii, also not observed. From this, one can assume that the damaging emission is dominated by field emission which is restricted to a limited range of phases ($\pm~\approx 20^{\circ}),$ and that  another mechanisms, such as space charge, spreads the beamlets out to a greater extent than this phase dependent effect. Figure~\ref{vsphase} shows the energies of the electrons on impact and indicates which phases are returned and which arrive on the next iris. 
Field emitted electrons $(\pm 25^{\circ})$ do not come back, but are focused to a local area on the opposite iris. Experimental observations~\cite{noremopen} of the time structure of dark current show that this dark current is concentrated at phases close to zero. It seems unlikely then that, even with an external axial magnetic field, significant currents of electrons will be returned to their source.

The electron energies for the 805~MHz cavity with axial rf fields of 25~MV/m are approximately 1~MeV. For a 201~MHz cavity, and the same gradient, they are of the order of 4~MeV. At these energies, the electrons  penetrate to significant depths in the Cu cavity walls. If Be is used, the penetration is even deeper. The relative surface heating and thus probability of melting and damage depends on the fraction of energy deposited in a surface layer, taking into account thermal conduction away from the deposition. A full calculation would include the time dependence of both deposition and conduction. As an approximation, the temperature rise is estimated from the deposition  in a depth corresponding to the thermal diffusion depth at the surface. 
\subsection{The Effects of Space Charge on the Transverse Distribution of Field Emitted Current}
Without an asperity and emission from a small area, the space charge forces give transverse momenta to emitted electrons causing the beamlet's radius to increase. As the beamlet increases in radius and the electrons are accelerated, the space charge forces drop and it can be shown that the induced \textsl{rms} transverse momentum is $\sigma_{p\perp}\propto \sqrt{I}.$ But if the electrons are emitted from the tip of an asperity then they will first be spread by the approximately spherically symmetric local electric fields, and the effect of the space charge is consequently modified.
 
\begin{figure}
\includegraphics[width=2.75in]{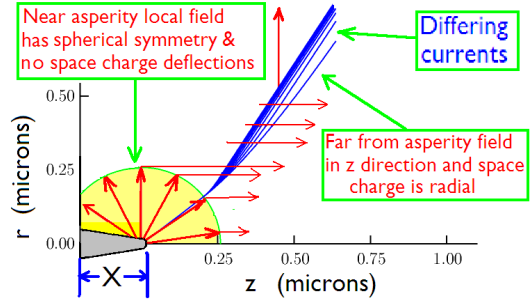}
\caption{\label{spacech}(Color) Schematic of approximate simulation of space charge effects on electrons emitted by an asperity.   }
\end{figure}

A simple simulation was performed (see Fig.~\ref{spacech}). The initial electric fields were assumed to have strength $\beta_{\text{FN}}\bm{E}$, and exact spherical symmetry out to a distance $X$. Beyond this distance, the fields were assumed to be perpendicular to the average surface with strength $\bm{E}$. No space charge effects are included in the spherically symmetric part. Radial space charge forces, inversely proportional to an average radius, are assumed beyond the distance $X$.
 We found that with  $X=0.2~\mu$m, the transverse momenta could be approximated by $\sigma_{p\perp}\propto {I^j},$ 
 but with the power $j=0.3$ instead of 0.5 as in the simple case without asperity. 
\begin{figure}
\includegraphics[bb=10 500 530 765,clip,width=3.in]{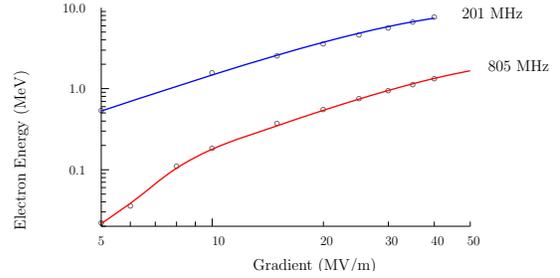}
\caption{\label{energy}(Color) The simulated final electron energy ${\cal E}_e$ as a function of axial rf gradient for (red) a 805~MHz pillbox cavity, and (blue) a 201~MHz cavity.}
\end{figure}
If $X$ was smaller, then the best fit exponent $j$  was found to increase in value. Since neither the asperity height, nor its shape, are known, it is reasonable to treat the exponent $j$ as an unknown that is fitted to the experimental data. In a better simulation, this fit would give us information on the asperity dimensions, but this model is too simple for this at this stage. At  distances from the source large compared with $X$, space charge becomes negligible, but  the transverse momentum is focused by the axial magnetic field, giving a beamlet with an rms radial size $\sigma_r \propto \frac{I^j}{B}$ where $B$ is the axial magnetic field. The power per unit area $W$ of the electron beamlet hitting the  opposing surface is given by 
\begin{equation}
W= \frac{I{\cal E}_e}{\pi\sigma_r^2}~\propto \frac{I^{(1-2j)}~{\cal E}_e~B^2}{\pi }
\end{equation}
where ${\cal E}_e$ is the final electron energy in $(\text{MeV})$ determined from CAVEL~\cite{cavel} simulations. ${\cal E}_e$ is plotted in Fig.~\ref{energy} as a function of the axial rf gradient, for 1) 805~MHz pillbox cavity~\cite{norempill}, and 2) 201~MHz cavity designed for the MICE cooling experiment~\cite{mice}.
\subsection{Fraction of Energy Deposited in the Thermal Diffusion Depth}
The  thermal diffusion length $\delta$ corresponding to the rf field pulse duration $\tau (s)$  is, $\delta=10^{-2}\sqrt{D~\tau}$ 
where $D=\frac{K}{\rho C_s}$ is the thermal diffusion constant in (m), $K$ is the thermal conductivity and $C_s$ the specific heat. It is assumed that all the energy deposited in this diffusion depth is spread uniformly within that depth $\delta.$ 
The penetration depth $d$ in $(\mu m)$ of low energy electrons is approximately given by~\cite{penetration} 
\begin{equation}
d=0.0267 \frac{A}{\rho Z^{0.89}}{\cal E}_e^{1.67}
\end{equation}
where ${\cal E}_e$ is the electron incident energy in (keV), $\rho$ is the material density in $(\frac{g}{cm^3}),$ and  $Z$ and $A$ are the atomic number and atomic weight respectively.  

If ${\cal E}(x)$ is the energy of an electron with penetration depth $x$ and we define $Q$ as the fraction of the electron energy deposited in the thermal diffusion length $\delta$, then: 
\begin{itemize}
\item For electrons whose penetration depth is less than the diffusion depth ($d<\delta$): 
$\quad Q=1$.
\item For electrons whose penetration depth is greater than the diffusion depth, but   not much greater ($\delta<d< 10\times \delta$) then:
$\quad Q=\frac{{\cal E}_e-{\cal E}_e(d-\delta)} {{\cal E}_e}.$ 
\item For electrons with  penetration depth $d > 10\times \delta$ then: $Q=\frac{\frac{d{\cal E}_e}{dx}\delta}{{\cal E}_e}.$
\end{itemize}
\subsection{Dependency of Local Temperature Rise}
The surface temperature rises as a fraction of the melting temperature  is then
\begin{equation}
\frac{\Delta T}{T_{\text{m}}}~\propto W\left(\frac{\tau Q}{\delta~ \rho~ C_s~T_{\text{m}}}  \right)
\end{equation}
where  $\rho$ is the density, $C_s$ is the specific heat, $T_m$ is the melting temperature, and $\tau$ is the rf pulse length, taken to be 20 and $160~\mu s$ at 805 and 201~MHz respectively ($\tau\propto\lambda^{3/2}$). 
Parameters used for Cu and Be are given in Table~\ref{tb2}.
\begin{table*}
\caption{\label{tb2}Atomic and Nuclear Properties of Cu and Be}
\begin{ruledtabular}
\begin{tabular}{l|ccccccccc}
\hline
&Z&A&K&$\rho$& $C_s$ & D&$\delta$(805)&$\delta$(201)& $T_{\text{m}}$\\
&&&(W/cm$-^o$C)&(g/cm$^3$)&(J/g$-^o$C)&(cm$^2$/s)&($\mu$m)&($\mu$m)&($^o$C)\\
\hline
Cu&29&63.5& 4.01 & 8.96&0.385&1.16&48 &186&1085\\
Be&4&9.0& 2.18&1.85&1.825&0.81&40&155&1287\\
\end{tabular}
\end{ruledtabular}
\end{table*}
\subsection{ Fit to Field Dependence Data from Lab-G Pillbox and Predictions for 201~MHz} 
The above is a very approximate analysis. A full simulation of the problem is being pursued. The
 simulations were done with uniform magnetic field; tracks were only simulated from the single maximum gradient location; electron impacts were assumed at $90^{\circ}$ to the surface; the thermal diffusion calculation ignored the rise time shape and used  an  approximate calculation; both current scale and space charge strength were normalized to fit data.
However, if this simple approach can qualitatively fit the data, it should allow a qualitative extrapolation to 201~MHz and other materials. A fuller simulation should provide more quantitative results.  
The red curved line in Fig.~\ref{fit} shows the fit to the 805~MHz Cu cavity in Lab G breakdown data~\cite{norempill}; 
 the fitted value of the current exponent was $j=0.35.$ More data~\cite{july} has been taken since the cavity had suffered significant damage. These data  (not shown) lie at somewhat lower breakdown gradients and presumably correspond to a now higher value of the Fowler-Nordheim $\beta_{\text{FN}}$. But since
 $\beta_{\text{FN}}$,  has not been redetermined from dark current measurements of the damaged cavity,  this analysis with that data is not possible. The black line in the plot shows the dependence predicted by the twist model~\cite{noremopen}, which does not fit the data well. The dashed lines  show the calculated breakdown limits for Be. The horizontal red line indicates the gradient limit from the assumed model without magnetic fields, assuming the local field limit to be 7~GV/m.
\subsection{Extrapolation to 201~MHz}
Without an external magnetic field,  breakdown gradients have been observed to follow approximately a $\sqrt{f}$ behavior (see. Sec.~\ref{vacnofield}). Under the assumptions used here~\cite{noremmodel}, the local field at breakdown is  independent of the frequency; this implies that $\beta_{\text{FN}}$ decreases approximately  as $\frac{1}{\sqrt{f}}.$ 
Using this assumption, and using the appropriately modified diffusion depth, the predicted breakdown limits at 201~MHz, for  Cu (solid blue line) and Be (dashed blue line) are shown in Fig.~\ref{fit}. The predicted 201~MHz breakdown gradients are seen to be a factor of 2 to 2.5 below those for 805~MHz. This factor comes primarily from the higher expected $\beta_{\text{FN}}$, but also from the longer rf pulse duration and thus longer time to heat and melt the surfaces. 
At the higher gradients specified for Neutrino Factory phase rotation and cooling (15~MV/m at B$\le$ 3~T) are well above this prediction. 
\subsection{ Other Experimental Results Consistent with this Analysis}
\begin{figure}
\includegraphics[width=3.in]{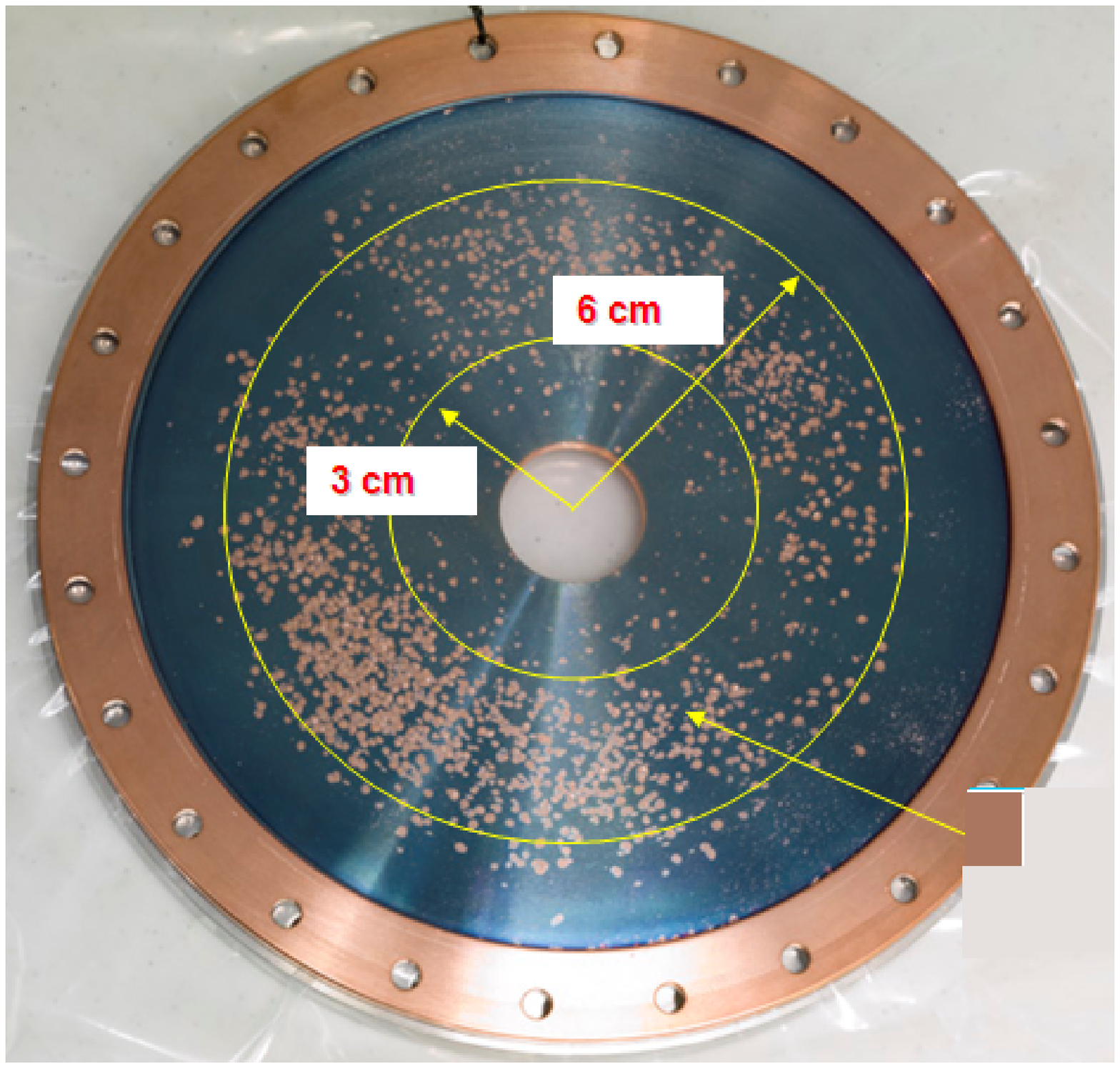}
\includegraphics[width=3.in]{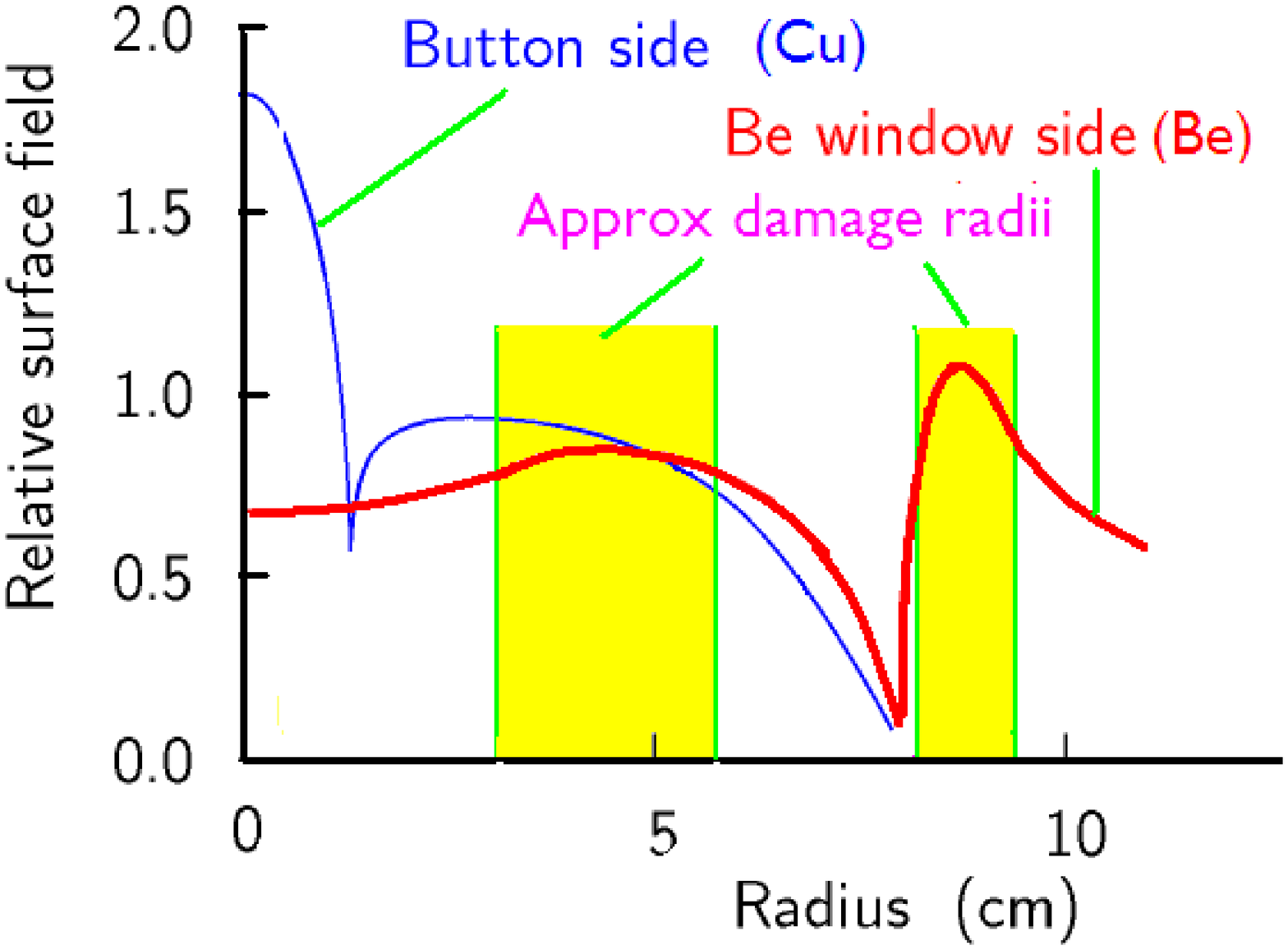}
\caption{\label{button}(Color) Top: a) Button support plate showing damage band between 3 and 6~cm radius. Bottom: b) Surface fields on Be (red) and Cu (blue) \textit{vs} radius, with bands showing where damage was concentrated. (Data from Ref.~\onlinecite{button})}
\end{figure}

The pillbox cavity used in the above experimental study has, more recently, been tested~\cite{button} with one of its two Be window replaced by a flat Cu plate with an easily replaceable central `button'. The button had dimensions such that the local field on the tip  was 1.7 times that on the outer Cu `iris'. The intent was to allow a study of magnetic field dependent breakdown as a function of materials, the assumption being that breakdown would occur at the high gradient on the tip of the button. The cavity was found to operate with gradients on the button significantly higher (by approximately a factor of 1.7)
 than had been observed without the button. It was noted however that this breakdown occurred with gradients on the iris, and on the flat support plate, that were essentially the same as those present without the button. This  suggested that breakdown was not occurring on the button, but rather at other locations that did not have the 1.7 times field enhancement: on the irises and/or the flat button support plate. When the cavity was later disassembled it was indeed found that was there little damage on the  button, where the field was maximum. But there was significant damage in a distinct band from 3 to 6~cm on the TiN coated Cu support plate (see Fig.~ \ref{button}a).
    
These observations are consistent with the breakdown mechanism described in this paper. SUPERFISH~\cite{superfish} calculations showed that the fields on the Be window opposite the button and support plate were maximal just in the band 3 to 6~cm, where the damage was observed (see Fig.~\ref{button}b). In this model, emission from the Cu button, falling on the Be is less liable to cause damage and breakdown, whereas the emission from the Be focused onto the Cu plate should cause damage just where the gradients were maximal on the Be.
 \section{ Possible Solutions, or Reductions, of these Problems}
There are several  possible approaches to these breakdown problems in a Muon Collider or Neutrino Factory.
\begin{itemize}
\item Redesign the phase rotation and cooling channels to use lower rf fields. This approach would clearly hurt performance, and, in addition, risks a slow deterioration of performance as occasional breakdown events continue to spray Cu around the cavity - as observed in the pillbox tests.
\item Use cooling lattices with high pressure hydrogen gas in the rf cavities. No degradation of rf performance has been observed in a small 805~MHz test cavity with axial magnetic fields and rf gradients similar to those in vacuum cavities. If the loading from beam induced electrons is not a problem, or is slowed by introducing gas impurities, then this solution should offer no loss of performance in early cooling stages. But to cool to very low emittances, it will probably require lattices with lower Courant Snyder $\beta$ at the absorber than can be achieved in the rf. In this case, the addition of hydrogen gas at the higher $\beta$s in the rf would cause unacceptable emittance growth. So for later cooling, high pressure hydrogen gas is probably not a solution.
\item Build cavities with exceptionally good surfaces so that the $\beta_{\text{FN}}$ is sufficiently low initially that no breakdown occurs. With Atomic Layer Deposition (ALD) this may be a realistic option~\cite{ald}. The fear would be that a single breakdown spoils the surface in such a way that there will be a cause of further breakdown and a conditioning that approaches the same gradient limits seen in a conventional cavity.
\item Design lattices with magnetic field shielded from the rf. The above prediction suggests that so long as the field is less than about 0.2~T, no adverse effects will be observed. Attempts to design such lattices have, unfortunately, shown significantly worse performance.
\item Design lattices  using multi-cell open cavities, with alternating  current coils in their irises (Fig.~\ref{tracks}). In this case, as in the original multi-cell open cavity tests, the focused electrons would be directed to low field regions in the cavity and would thus not initiate breakdowns. Nevertheless damage  done to those locations and molten Cu   ejected from such damage could cause eventual deterioration of performance. In addition, the use of open, instead of pill-box cavities implies lower acceleration for a given surface field.
\begin{figure}
\includegraphics[width=2.5in]{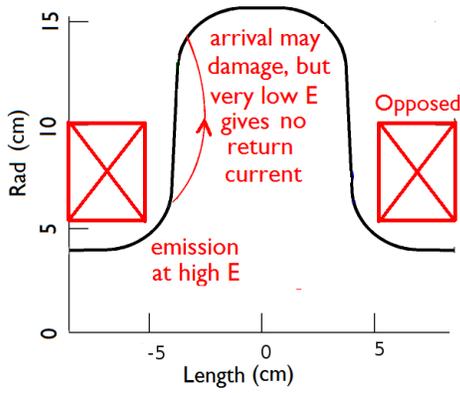}
\caption{\label{tracks}(Color) Open cavity with alternated solenoid coils in irises.     }
\end{figure}
\item Damage might be eliminated if cavities were designed such that all high electric gradient surfaces were parallel to the magnetic fields (Fig.~\ref{magins}). This could provide `magnetic insulation'~\cite{maginsul}. Dark current electrons would be constrained to move within short distances of the surfaces, would gain little energy, would cause no X-rays, and do no damage.
Possible difficulties might be: a) cavities so designed will not give optimum acceleration for given surface fields, and b) multipactoring might occur, now that the energies with which electrons do return to the surfaces are in the few hundred volt range where secondary emission is maximal. 
In addition, the use of open, instead of pill-box, cavities implies lower acceleration for given surface fields.
\begin{figure}
\includegraphics[bb=98 432 581 701,clip,width=3.in]{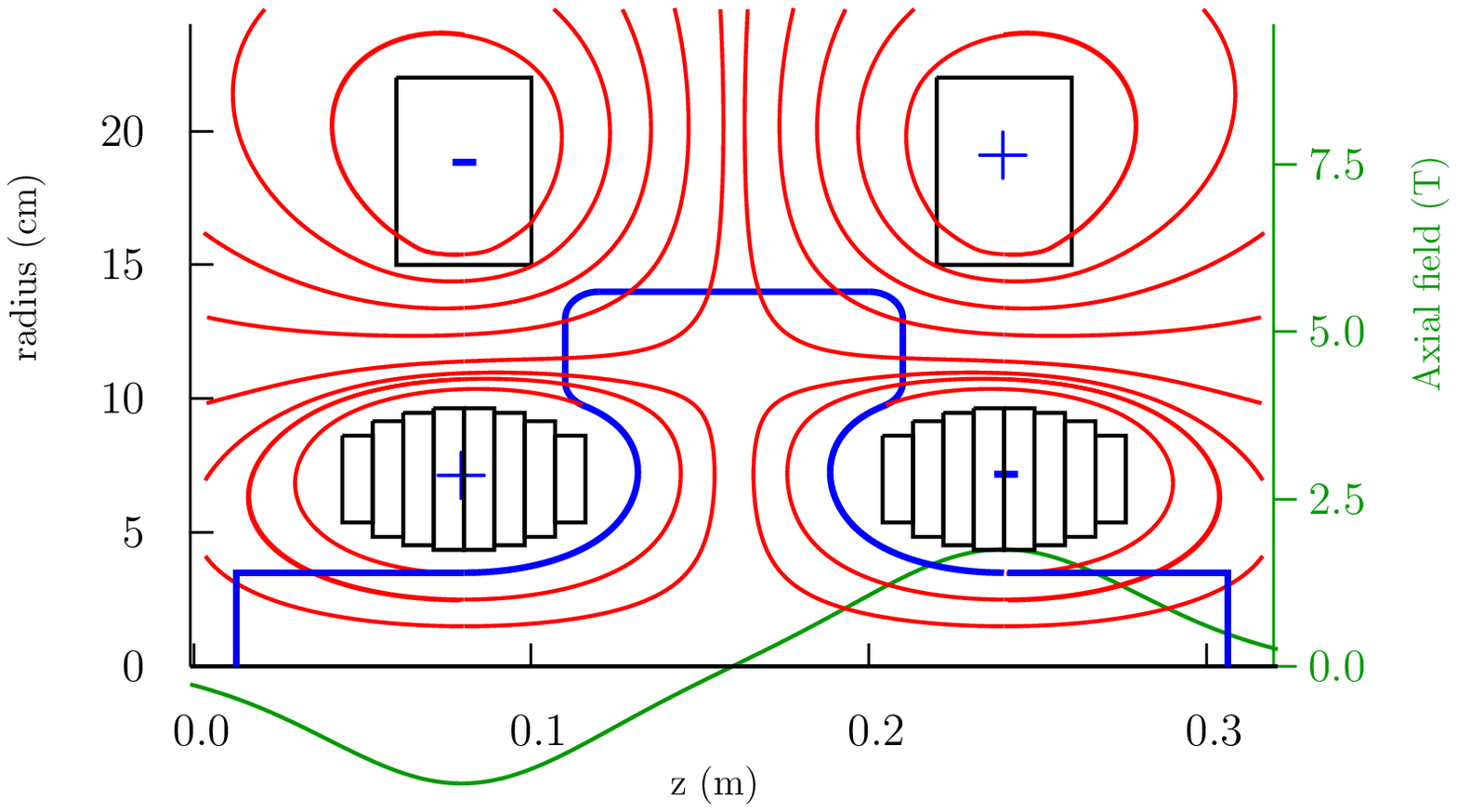}
\includegraphics[bb=20 485 464 727,clip,width=3.in]{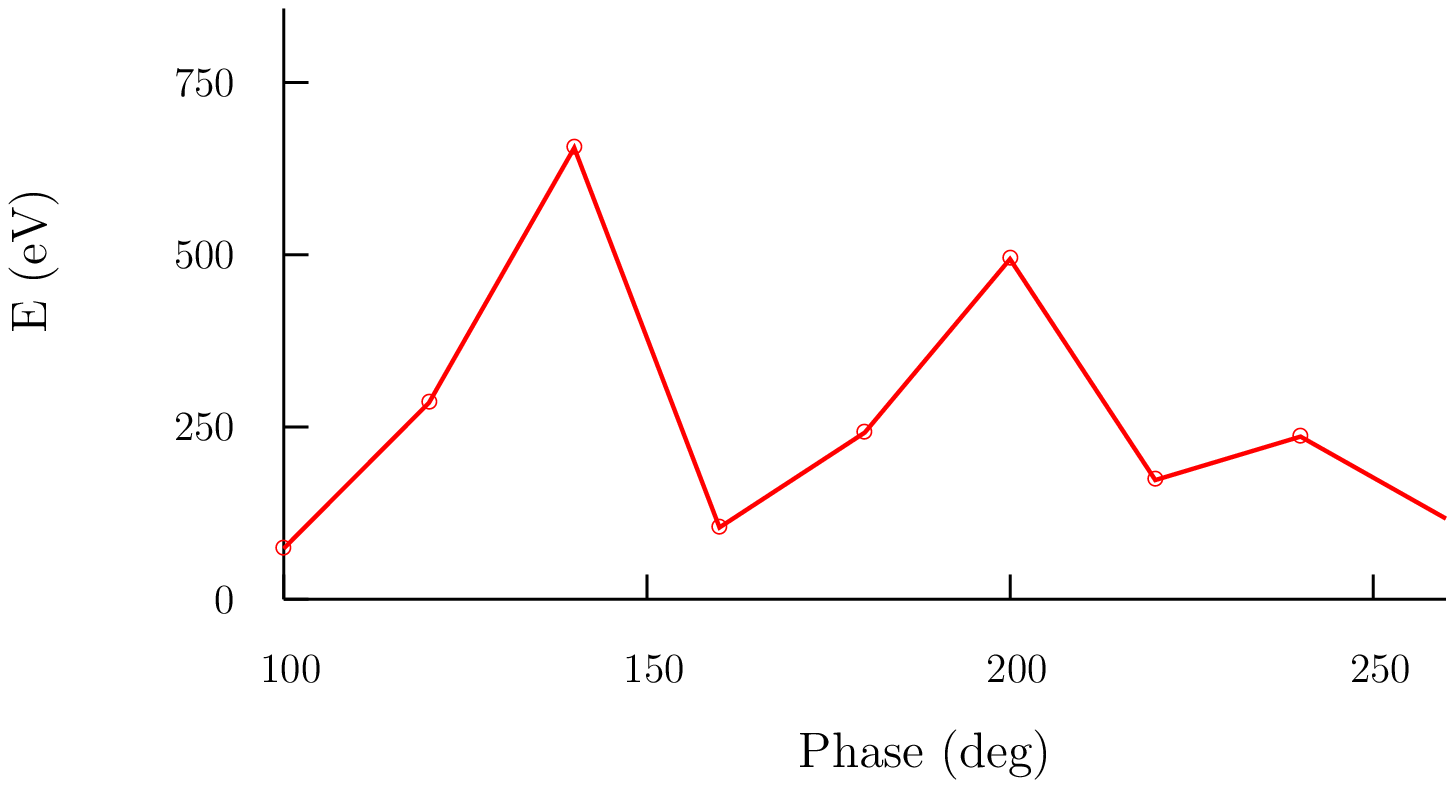}
\caption{\label{magins}(Color) Top: a) Magnetic field lines from coils in a cavity lattice, together with cavity shape that follows these field lines. Bottom: b) Energies of returning electrons as a function of their initial phase.   }
\end{figure}
\end{itemize}
\section{Experiments needed to study these problems}
Two critical experiments are already planned: 

1) The testing of the existing 201~MHz cavity in magnetic fields similar to those in current MICE and cooling designs. These experiments would use a MICE `Coupling Coil' when it is available.

2) Operating a high pressure hydrogen filled test cavity in a proton beam to study the possible breakdown and rf losses due to the ionization of the gas by the beam.

Whatever the results of these tests, further experiments will likely be needed to study the observed problems at 805~MHz and test possible solutions. If the tests of the 201~MHz cavity in magnetic fields show problems also at that frequency, then further 805~MHz tests would be needed to explore solutions for these lower frequencies. Eventually, however, tests of any proposed solutions would have to be done at 201~MHz.   

If with the proton beam the rf losses in the 805~MHz test cavity are sufficiently low, then further tests of high pressure gas filled cavities will be needed. In particular, cavities must be tested with more stored energy and with a similar shape to those needed. Thin windows must be designed, safety problems must be addressed and beams with time structure and intensity nearer to those in the applications should be be employed.
\subsection{Experiments with a Simple Vacuum Pillbox Cavity}
\begin{figure}
\includegraphics[width=3.in]{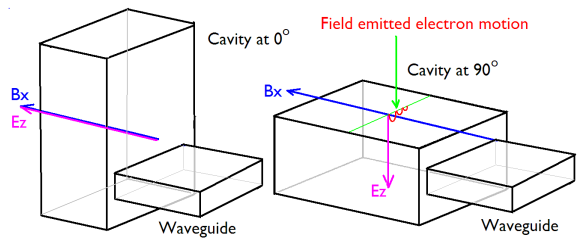}
\caption{\label{square}(Color) Simple pillbox cavity with mounting in two orientations within the lab G solenoid.}
\end{figure}
The `pillbox' cavities that have so far been tested have relatively complex shapes, and the simulations of their performance with fields in differing directions are complex. It would thus be desirable to test a simple pillbox shaped cavity whose performance would be far easier to simulate. The cavity should be designed so that it can be mounted in the center of the of the Lab G magnet in at least two orientations: with its axis parallel to the solenoids and perpendicular to that angle. Ideally it should be possible to test it at other angles as well. A square cavity~\cite{moretti1} (Fig.~\ref{square}) would meet this requirements, would be easy to build and easy to arrange with coupling ports to the waveguide in either of two locations to meet the two angle requirement.   
Such a simple cavity design would also be a good test vehicle for testing surface treatments including Atomic Layer Deposition (ALD). Since the design is so simple, several versions could be made to compare their performances.

\begin{figure}
\includegraphics[width=2.5in]{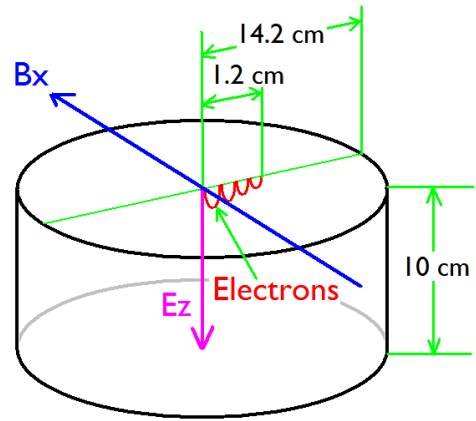}
\caption{\label{simple}(Color) A schematic illustration of an emitted electron's orbit when the magnetic field is parallel with the emitting surface.}
\end{figure}

As illustrated in Fig.~\ref{simple}, CAVEL simulations of a cavity with its axis perpendicular to the solenoid axis shows that the electrons are constrained to lie within a very small distance from the surface and gain little energy.  Such a  cavity would  test whether multipactoring is a serious problem. 
In  addition, if such a cavity were tested with the emission surface at a small angle to the field then the electrons could gain some energy, but would end up on the cylindrical outer surfaces where there would be no electric field. This would test the configuration of the coil-in-iris solution without the shape modified to achieve magnetic insulation. Surface damage might be expected, but it should not cause breakdown.
\subsection{A Single Cell Experiment with  `Magnetic Insulation'}
\begin{figure}
\includegraphics[width=3.in]{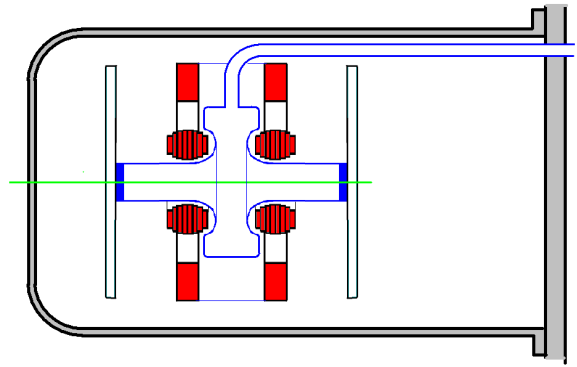}
\caption{\label{single}(Color) Single cell with magnetic insulation.}
\end{figure}
In this experiment (Fig.~\ref{single}) superconducting coils would be mounted on either side of a cavity whose shape is such that the  magneric fields are strictly parallel with the high gradient surfaces. The two coils close to the open pipe are powered in opposite directions. A second pair of coils is mounted outside the inner coils and powered with currents in the opposite direction to those they surround. The coils would be separately powered so that the sensitivity to deviations from the insulated condition can be studied.

The rf cavity would be operated at liquid nitrogen temperatures in order to minimize the radiation falling on the superconducting coils that will be very close to its surfaces. There would be a vacuum both outside the rf cavity (for thermal insulation) and inside the cavity, although the quality of the latter should be higher, and would be separately monitored. Super-insulation and an outer liquid nitrogen shield are needed but not shown. Tuning of the cavity would be provided by axially squeezing or stretching the cavity.   
   A variant of this experiment would have Be windows, making it into more of a pillbox design and raising the accelerating field. This would not provide such complete magnetic insulation but, because of the special properties of the Be might still have acceptable performance.
\subsection{A Multi-purpose Test Stand for these Experiments}
It is proposed that this and the following experiments would be carried out on a multi-experiment  test stand. To allow ease of assembly, mounting of instrumentation, and making changes, all connections (rf, cryogens, magnet and instrumentation leads etc.) would be brought in through a single support plate (on the right side in the illustration). The vacuum container would be in the form of a dome that joins to the support plate with a single flange, and can thus be easily removed without disturbing the connections. 
\subsection{A Multi-cell Experiment with  `Magnetic Insulation'}
\begin{figure}[!th]
\includegraphics[width=3.in]{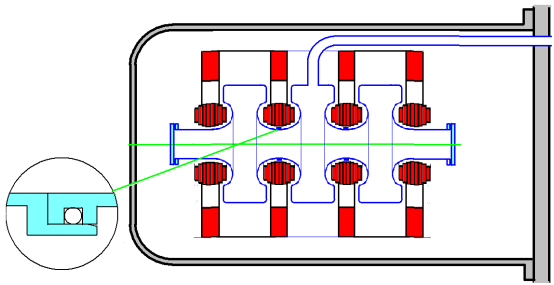}
\caption{\label{multi}(Color) Multi-cell cavity with magnetic insulation.}
\end{figure}
The next experiment would be of a multi-cell magnetically insulated cavity (Fig.~\ref{multi}).  This would test the magnetic insulation in a geometry similar to that required in the 201~MHz acceleration for phase rotation or early cooling. In addition, it would address the problem of joining cavities inside the bore of the solenoids. This joint does not need to carry much rf current, nor does it need to hold high vacuum (since there is vacuum on both sides). The method requiring simple axial pressure is shown in the figure, but other ideas may be explored.

It should be noted that this experiment, and the previous one, test a model of the 201~MHz lattices that would be used in the phase rotation, Neutrino Factory cooling, and the early cooling in a Muon Collider. 
\subsection{Testing of Components for later 6D Cooling}
\begin{figure}[!bh]
\includegraphics[width=3.in]{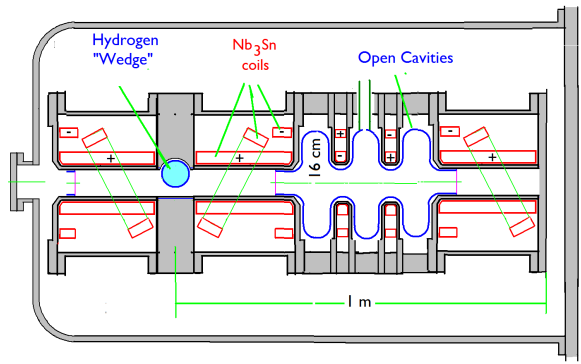}
\caption{\label{later}(Color) Lattice for later cooling to lower emittances.   }
\end{figure}

Later 6D cooling will require compact higher field (10-15~T) solenoids in order to focus the beams to lower $\beta$'s and thus cool to lower emittances. Test coils could conveniently be tested on the proposed test stand, since it would allow for easy mounting, cooling and testing with easily accessible instrumentation. Later, these coils would be used in conjunction with 805~MHz cavities (see Fig.~\ref{later}).
The first stage of the 10-20~T solenoid development could appropriately be tested in the same test stand discussed above. Subsequently, the rf cavities would be included. The final step would be to add the required hydrogen absorber. The details of the lattice, whether it is a snake~\cite{alexahin1} or Guggenheim~\cite{Gug} configuration for instance, would not change the nature of the components needing test. Figure~\ref{later} is thus not meant as a definitive design of the required lattice, but rather as a test of the kind of components needed; the cavity shapes shown are not those with true magnetic insulation. The true shapes with their coils are yet to be determined. Figure~\ref{later} is thus intended only as an illustration of the direction of the required R\& D.
\subsection{Testing of Magnetic Field Solutions at 201~MHz}
The experiments already planned with the 201~MHz cavity in several Tesla fields could show satisfactory performance. If however this is not the case, then whatever solutions are proposed on the basis of the 805~MHz tests will need to be demonstrated at 201~MHz.
\section{Conclusions}
The main results and predictions of the model presented in this paper can be summarized as:
\begin{itemize}
\item A review of models for breakdown without external magnetic fields suggests that the initiation of breakdown is best explaned by ohmic heating due to field emission current at asperities.  The dependence of breakdown on frequency, pulse length and cavity materials can be explained by the competing processes of asperity destruction and the creation of new asperities.
\item We have proposed a model for damage in rf cavities operated in significant axial magnetic fields. The model fits the existing data reasonably well.
\item The model also fits some otherwise surprising results from a pillbox cavity with a Be window facing a Cu plate with a central button. 
\item The model predicts relatively low gradient breakdowns at  201~MHz in magnetic fields.
These predicted breakdowns occur at significantly lower gradients than the operating gradients specified for the phase rotation and initial cooling in the ISS~\cite{iss} Neutrino Factory  and in a Muon Collider.
\item Methods to address these problems are discussed, including the use of `magnetically insulated rf'. An experimental program to study this concept is outlined.
\end{itemize}
\begin{acknowledgments}
We would like to thank J.~Norem, A.~Moretti and A.~Bross for many discussions and sharing the experimental data. This work has been supported by U.S. Department of Energy under contracts AC02-98CH10886 and DE-AC02-76CH03000.
\end{acknowledgments}

\end{document}